\documentclass{appolb} 
\usepackage{epsfig} 
\usepackage{amssymb}
\usepackage{color}
\begin{document}

\pagestyle{plain}

\title{Multifractal Model of Asset Returns versus real stock market 
dynamics}

\author{P.~O\'swi\c ecimka$^1$, J.~Kwapie\'n$^1$, S.~Dro\.zd\.z$^{1,2}$, 
A. Z.~G\.orski$^1$, R. Rak$^2$ \address{$^1$Institute of Nuclear Physics, 
Polish Academy of Sciences, \\ PL--31-342 Krak\'ow, Poland\\ $^2$Institute 
of Physics, University of Rzesz\'ow, PL--35-310 Rzesz\'ow, Poland}}

\maketitle

\begin{abstract} 

There is more and more empirical evidence that multifractality constitutes 
another and perhaps the most significant financial stylized fact. 
A realistic model of the financial dynamics should therefore incorporate 
this effect. The most promising in this respect is the Multifractal Model 
of Asset Returns (MMAR) introduced by Mandelbrot {\it et al.}~\cite{mandelbrot1997b} in which 
multifractality is carried by time deformation. In our study we focus on 
the Lux extension to MMAR and empirical data from Warsaw Stock Exchange.
We show that this model is able to reproduce relevant aspects of the real stock market dynamics.

\end{abstract}

\section{Introduction} 

Empirical data collected from the stock market has extremely complicated
structure. Apparently, it can seem to be white noise without correlations
and with complete disorder, but, by investigating it deeper by means of
sophisticated methods, one discovers important non-random features and an
amazing hierarchy. Amongst these key properties (the so-called stylized
facts) are the fat tails of p.d.f. of the logarithmic returns, the long
range correlations in volatility, the leverage effect and some nontrivial
fractal characteristics of data ~\cite{gopikrishnan1999a,eisler2004,matia2003,drozdz2003,kwapien2005a,oswiecimka2005a}. Especially the fractal aspects seem to be particularly important due to their ability to describe complex systems in
a relatively simple way. Scaling properties, which are fundamental in the
fractal formalism, manifest themselves as a linear plot of a given
quantity on a log-log scale and the systems exhibiting such behaviour are
called scale-free. As many real-world signals, the financial data cannot
be described as a single fractal object but rather as a structure
comprising a whole family of interwoven fractals forming a multifractal.
Thus, description of this kind of data requires idetifying all of its
fractal components. The multifractal formalism was originally proposed in
ref.~\cite{halsey86} where the singularity spectrum $f(\alpha)$ was
introduced as a tool characterizing a multifractal. This formalism allowed
researchers to discover multifractality in almost all fields of science
and has gradually become a powerful method quantifying time series coming
from biology, physics, economics, technical science and many others \cite{ivanov1999,stanley1999b,bunde1991}.

Having identified the fractal properties of data, the next natural step of
the analysis is to simulate a process capable of reproducing these
properties under control. As regards the financial world, the two
following models have been proposed. The Mandelbrot's Multifractal Model
of Assets Returns (MMAR)~\cite{mandelbrot1997b,calvet1997,fisher1997} approximates evolution of the returns
with a compound process involving multifractal time and a monofractal
fractional Brownian motion. In contrast, the Multifractal Random Walk
(MRW)~\cite{bacry01,muzy2000} can be viewed as a continuous interpolation of a
discrete multifractal cascade. What is particularly important in this
context is that multifractal features are not reproduced in such popular
models like the family of ARCH processes.

In our paper we refer only to MMAR model and, more specifically, to one of
its extensions that we shall call the Lux model~\cite{lux2003b}. Unlike the static
original MMAR model, it incorporates an iterative mechanism that mimics
effects of the generating cascade for each consecutive point without
limiting the length of a signal. An advantage of this innovation, that
looks attractive especially from the perspective of financial engineering,
is its forecasting power~\cite{lux2004a}. We show flexibility of the Lux model while an
analyzed process changes dynamics. Inspecting the evolution of the model's
parameters can help us interpreting the multifractal properties of
empirical data.

\section{Multifractal formalism}

Let us denote the increments of a stationary process $X(t)$ by \cite{mandelbrot1997b}
\begin{equation} 
X(t,\Delta t)=X(t+\Delta t)-X(t) .
\end{equation} 
Multifractality of this process can be defined by a non-uniform scaling 
of the q-order correlation function $X(t,\Delta t)$ 
\begin{equation} 
E[X(t,\Delta t)^q] \sim \Delta t^{\tau(q)+1} ,
\end{equation} 
where $\tau(q)$ denotes the scaling exponent. If $\tau(q)$ nonlinearly 
depends on $q$ we say that $X(t,\Delta t)$ possesses a multifractal 
character. In a case of linear behaviour of $\tau(q)$ such a process is 
called monofractal. An example of the monofractal dynamics is fractional 
Brownian motion which has no correlation (its classic version) or is only 
linearly correlated. $\tau(q)$ can be related to the singularity spectrum 
$f(\alpha)$ by applying the Legendre transformation~\cite{muzy1994}
\begin{equation} 
\bigg\{ 
\begin{array}{l} q = df/d\alpha \\ 
\tau(q)=q\alpha-f(\alpha) \ .
\end{array} 
\end{equation} 
For the Brownian motion the singularity spectrum consists of a single 
point localized at $\alpha=H$ ($H$ is the so-called Hurst exponent) and $f(\alpha)=1$, indicating a strong 
linear correlation (positive one for $H>0.5$ and negative one for $H<0.5$) 
or no correlation ($H=0.5$). For a signal with a multifractal structure 
the $f(\alpha)$ shape resembles an inverted parabola. The strength of 
multifractality can be quantified in terms of the $f(\alpha)$ curve width 
\begin{equation} 
\Delta \alpha= \alpha_{max} - \alpha_{min} .
\end{equation} 
The bigger $\Delta\alpha$ the richer multifractal (we can also say 
the more ``convoluted'' fractals). Multifractality of a signal can be 
detected by using one of the modern methods like Multifractal Detrended 
Fluctuation Analysis (MFDFA)~\cite{kantelhardt2002} or Wavelet Transform Modulus Maxima (WTMM)~\cite{muzy1994}. 
Here we restrict the analysis to MFDFA which in our opinion gives more 
reliable results~\cite{oswiecimka06}.

\section{Nonlinear dynamics of WIG20 }

\begin{figure}[t] 
\begin{center} 
\includegraphics[width=0.7\textwidth,height=0.6\textwidth]{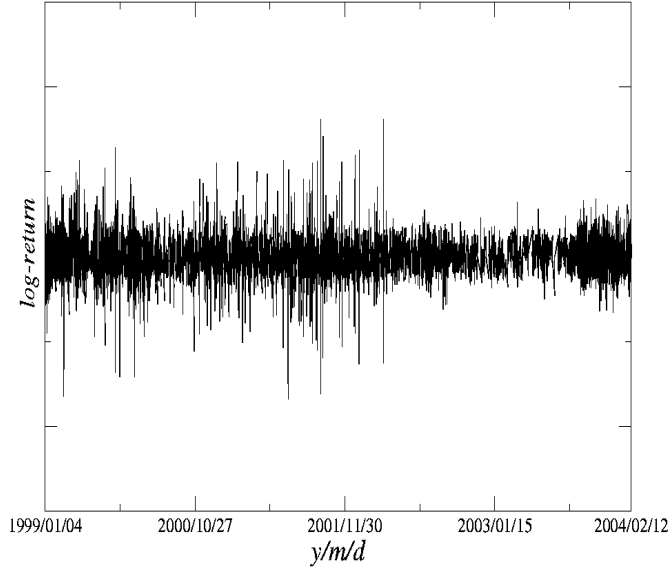}
\caption{Time series of WIG20 log-returns covering the analyzed period.} 
\label{WIG20 fluctuation} 
\end{center}
\end{figure}

\begin{figure}[t] 
\begin{center} 
\includegraphics[width=0.7\textwidth,height=0.6\textwidth]{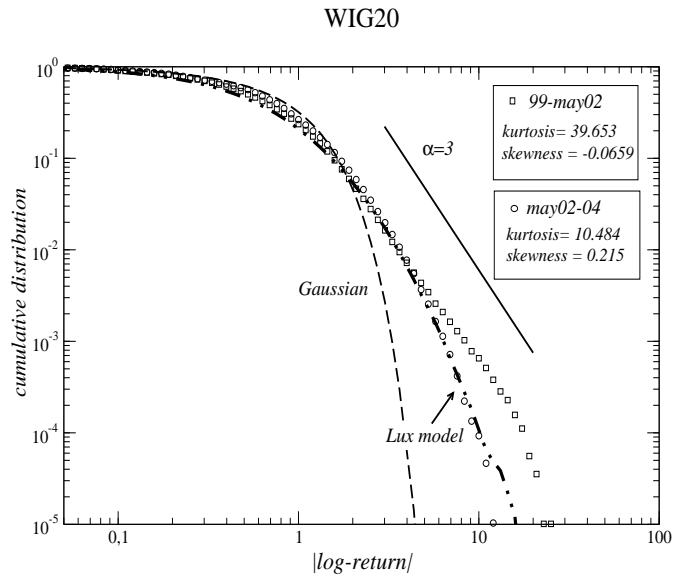}
\caption{Cumulative distribution of WIG20 log-returns for periods 
1999/01$-$2002/04 (squares) and 2002/05$-$2004/12 (circles) confronted 
against c.d.f.s of a Gaussian (dashed line) and of a signal generated 
according to Lux model (dash-dotted line).}
\label{WIG20 distribution} 
\end{center} 
\end{figure}

\begin{figure}[t]
\begin{center} 
\includegraphics[width=0.7\textwidth,height=0.6\textwidth]{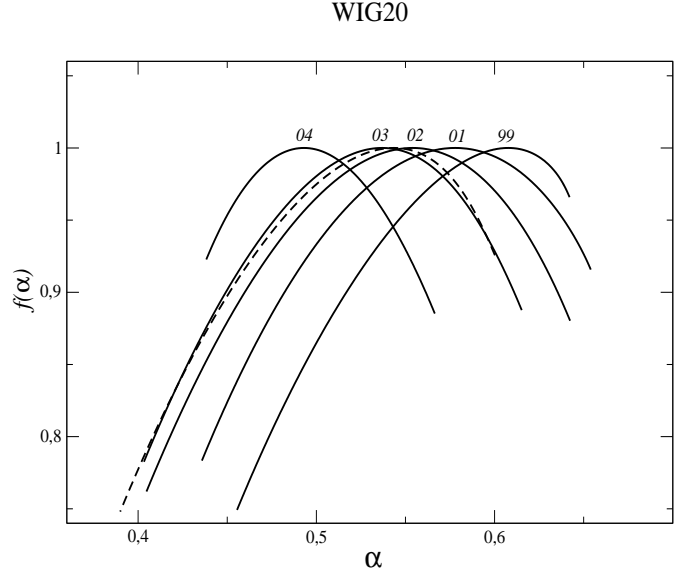}
\caption{Singularity spectra of WIG20 log-returns splitted into annual 
intervals; two-digit numbers denote the corresponding year except for 2000 denoted by dashed line.}
\label{WIG20 falfa}
\end{center}
\end{figure}

Recent studies show that fluctuation distribution of a market index cannot 
be well approximated by a Gaussian~\cite{gopikrishnan1999a}. Probability density function 
of the returns is characterized by the so-called fat tails which in most 
cases obey an inverse cubic power law if expressed as 
c.d.f.~\cite{gabaix03}. However, in the case of an emerging market we 
cannot {\it a priori} exclude a different behaviour. In the beginning of 
our study we first verify if the stock price fluctuations on the Warsaw 
Stock Exchange (WSE) obey the same law as larger developed 
markets. Calculations were carried out on high frequency data of 
the WSE blue chips WIG20 index sampled with 1 minute resolution over the 
period 1999-2004. All the overnight returns were removed as they are 
contaminated by some spurious artificial effects. The WIG20 fluctuations 
are shown in Figure \ref{WIG20 fluctuation}.
What we can easily notice is that after the end of April 2002 the wildest
fluctuations seem to be suppressed and the signal becomes ``milder''. 
Thus, it is natural to divide our signal into the two following periods 
and consider them separately: the first one from January 1999 to April 
2002 (237304 points) and the second one from May 2002 to December 2004 
(237268 points). Cumulative distributions for both time series are 
presented in Figure \ref{WIG20 distribution}.
In average they remain consistent with the inverse power cubic law. For the first period we actually have $\alpha=2.8$ but only for a very narrow range of plot, while for the second period a larger deviation 
is visible ($\alpha=4.2$). Consistently with the above 
conclusion from Figure \ref{WIG20 fluctuation}, we see a difference in 
scaling properties between the periods. From a potential investor's 
point of view, an even more important information is stored in 
temporal correlations of the analyzed signal. This information can be 
derived from its multifractal properties that can be a result of both the 
wide fluctuations distribution and the correlations in higher moments.
It is instructive to calculate the $f(\alpha)$ spectra for each year 
individually and to identify changes in the signal evolution. The 
corresponding spectra are shown in Figure \ref{WIG20 falfa}. 
The most interesting feature of this Figure is that the $f(\alpha)$ maxima 
systematically shift towards lower $\alpha$'s: from $\alpha =0.61$ 
in 1999 to $\alpha =0.48$ in 2004), with the exception of 2000 
(dashed line), suggesting a gradual transition from a strong persistence 
(1999) to a weak antipersistence (2004). Almost all the spectra are wide 
($\Delta\alpha > 0.2$) and can be interpreted as a manifestation of strong 
multifractality and rich dynamics. The narrowest curve is for 2004 
($\Delta\alpha = 0.11$) so we can regard it as the least interesting one 
from the multifractal point of view.

\section{Multifractactal Model of Asset Returns}

According to MMAR the logarithmic price $P(t)$ is assumed to follow a
compound process consisting of a fractional Brownian motion $B_H(t)$ and 
a time $\theta(t)$:
\begin{equation}
P(t)=B_H(\theta(t)) .
\end{equation}
Here $B_H$ represents a monofractal process which is a sum of random 
variables sampled by c.d.f. of a multifractal measure. Both $B_H$ and 
$\theta(t)$ are independent. A crucial role in the considered process 
plays the virtual trading time which can be interpreted as a 
deformation of the homogeneous clock-time or as a local volatility 
corresponding to faster or slower trading. The linear correlation of 
$P(t)$ depends on the Hurst exponent $H$ fully characterizing the Brownian 
motion, whereas the multifractal properties are generated by a 
multiplicative cascade. It has to be noted that in the original formalism 
of ref.~\cite{mandelbrot1997b} the whole cascade is generated globally at the same 
moment for each level $k$. However, for the sake of prediction we need an 
iterative procedure that is able to differentiate past and future events.
This is the rationale behind the application of a multiplicative measure 
proposed by Lux~\cite{lux2003b}.

Instead of $\theta$ it is better to consider its increments $\theta'(t)$ 
expressed by~\cite{eisler2004}
\begin{equation}
\theta'(t)=2^k\prod_{i=1}^{k} m_i(t) ,
\end{equation}
where $2^k$ is a normalizing factor and $m_i$ is a random multiplier taken 
from the log-normal distribution in accordance with the formula
\begin{equation}
m_{t+1}^{i}=\bigg\{
\begin{array}{l} \exp (N(-\lambda \ln 2,2(\lambda -1)\ln 2))\\ 
m_t^{(i)} ,
\end{array}
\label{lognormal}
\end{equation} 
where $i=1,...,k$ and $N(\mu,\sigma^2)$ denotes the conventional normal distribution. The upper option is taken either with the probability $2^{-(k-i)}$
or if for any preceding $i$ this option has already been chosen. Otherwise 
the multiplier remains the same as for previous $t$. We can imitate in this 
way the structure of a binary cascade and, on average, preserve its 
essential features. Based on this construction we see that in order to 
describe the multifractal properties of such a cascade we need only one 
parameter $\lambda$. Theoretical multifractal spectrum is then given by
\begin{equation}
f(\alpha )=1-\frac{(\alpha-\lambda)^2}{4(\lambda-1)} .
\label{log normal cascade falfa}
\end{equation}
This formula implies that $f(\alpha)$ is symmetric and has a maximum 
localized at $\alpha=\lambda$. Under the above formalism a return can be 
viewed as a composition of local volatility and white noise~\cite{lux2003b}
\begin{equation}
x(t)=\sqrt{\theta'(t)}\sigma N(0,1)=\sqrt{2^k\prod_i^{k} m_i(t)}\sigma 
N(0,1) .
\end{equation}

\section{WIG20 modelling}

We try to mimic evolution of WIG20 based on the Lux extension to MMAR and
observe how the changes in WIG20 dynamics influence the parameter
$\lambda$ of the model. This gives us information about the model
flexibility and its ability to simulate a highly nonstationary process
like the WIG20 returns. Since fitting the model with a log-normal
multiplier distribution requires to estimate only the parameter $\lambda$
defining the distribution of Eq.~(\ref{lognormal}), it can be derived by
using the relation~\cite{calvet1997}
\begin{equation} 
f_P(\alpha)=f_\theta(\alpha /H)
\label{pricetime}
\end{equation}
and Eq.~(\ref{log normal cascade falfa}). First we consider the simplest 
case of $H=0.5$. The singularity spectrum can be estimated by means of 
MFDFA with polynomial order 2~\cite{oswiecimka2005a}. Figure \ref{WIG20 distribution} 
shows c.d.f. of a time series generated according to the Lux model 
(dash-dotted line). This distribution has a somewhat thinner tail than the inverse 
cubic power law, but it is similar to c.d.f. for WIG20 from the period 
2002/05$-$2004/12. In this case the model appropriately reconstructs the 
dynamics including large fluctuations.

\begin{figure}[t]
\begin{center} 
\includegraphics[width=0.8\textwidth,height=0.55\textwidth]{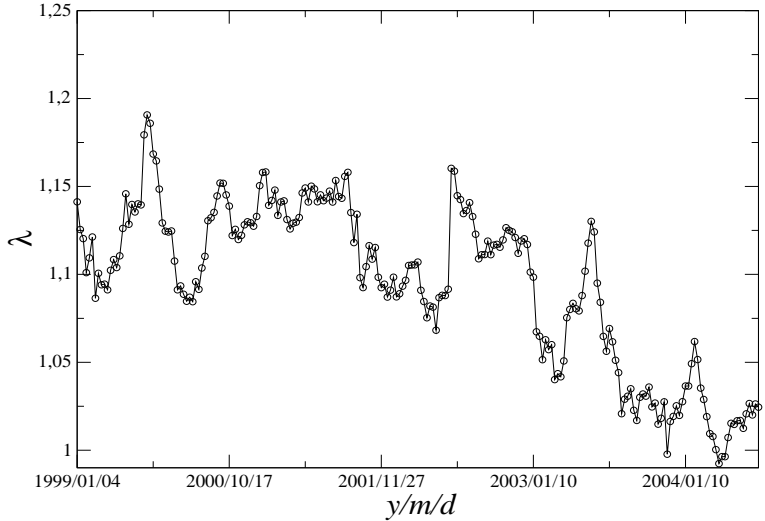} 
\caption{$\lambda$ calculated in a moving window as a function of time. 
Open symbols denote position of the first point for each window position.}
\label{lambda_co_1000} 
\end{center}
\end{figure}

The evolution of $\lambda$ is analyzed by using a moving window of length 
20000 data points shifted by 2000 points. Such a window ensures that we 
obtain statistically reliable results. From Figure \ref{lambda_co_1000} we 
see that $\lambda$ strongly fluctuates over the analyzed period. 
Nevertheless, starting from the end of 2001 it forms a clearly decreasing 
trend. 
The smallest values of $\lambda$ correspond to 2004 and even drop below 1. 
In the multifractal formalism $\lambda$ refers to the structure 
complexity; the larger $\lambda$ the more complicated signal and more 
complex fractal. It stems from this that from 2002 to 2004 the complexity 
of the Polish stock market gradually decreased despite the existence of 
some bigger fluctuations. This result remains in a perfect agreement with 
the indications of the $f(\alpha)$ spectra (Figure \ref{WIG20 falfa}).

\begin{figure}[t]
\begin{center} 
\includegraphics[width=0.8\textwidth,height=0.55\textwidth]{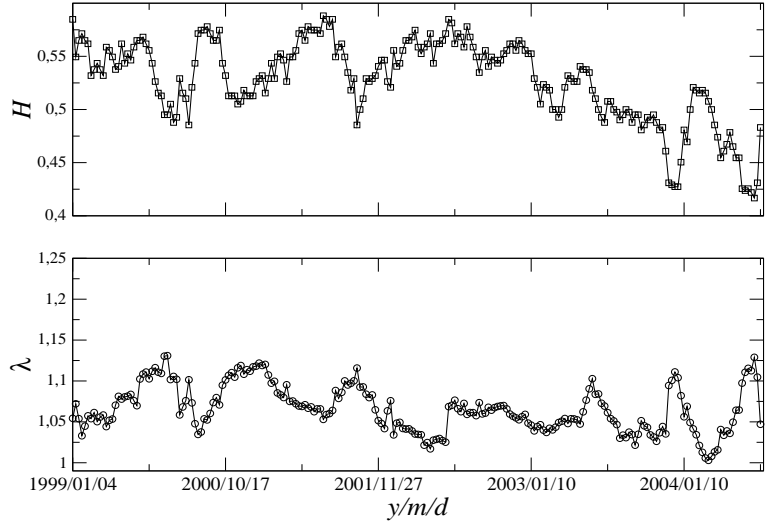}
\caption{Simultaneous temporal evolution of Hurst exponent $H$ (top) and 
parameter $\lambda$ (bottom). Open symbols denote position of the first 
point for each window position.}
\label{lambda_co_1000_i_H}
\end{center}
\end{figure}

Although fixing $H$ value at 0.5 can give approximate results for shorter 
signals (e.g. of length of the window applied before), in general it is 
recommended to treat $H$ as a free parameter which has to be estimated 
from the data because it may influence the forecasted volatility \cite{carbone2004}. To 
accomplish this task we have to use the relation between $f(\alpha)$ 
spectra for the price and the time processes (Eq.~(\ref{pricetime}))
together with the following condition~\cite{fisher1997}:
\begin{equation}
\tau_P(1/H)=0 .
\end{equation}
where $\tau_P$ stands for the scaling exponent of the price process. 
The so-estimated Hurst exponent $H$ can be used in the calculation of 
$\lambda$. Results for the windowed WIG20 are collected in Figure 
\ref{lambda_co_1000_i_H}. 
For the first three-year-long interval the Hurst exponent indicate 
persistent behaviour of the index, while starting from 2002 the value of 
$H$ almost monotonically decreases below 0.5 suggesting a transition to 
the antipersistent regime. Again, this resembles the conclusion made from 
Figure \ref{WIG20 falfa} but now the evidence is even more convincing.
As regards $\lambda$, we see that now its fluctuations are smaller and 
there is no clear decreasing trend. Therefore, in the present case of 
variable $H$ this quantity absorbs the information about changes in WIG20 
dynamics.

\section{Conclusions} 

The study presented in this paper reveals several interesting facts 
about dynamics of the Polish stock market. First, we see that the WIG20
fluctuation magnitude changes in the first half of year 2002. This 
suppression of the largest fluctuations is also visible in c.d.f. of 
the log-returns. Moreover, we identify this nonstationarity of WIG20 in evolution of the parameter 
$\lambda$ of the Lux model. These observations indicates that this model 
is able to reproduce at least some of the financial data characteristics
and to help one with detecting evolution phases with different properties.
In the last step we include in our analysis the empirical estimation of 
the Hurst exponent, a parameter which is necessary for the forecasting 
purpose. As our results document, value of this parameter largely 
determines the multifractal properties of the WIG20 evolution, leaving 
$\lambda$ only as an auxilliary measure.

\end{document}